\documentclass[12pt]{iopart}
\newcommand{\beq}{\begin{equation}}
\newcommand{\eeq}{\end{equation}}
\newcommand{\beqa}{\begin{eqnarray}}
\newcommand{\eeqa}{\end{eqnarray}}
\newcommand{\ba}{\begin{array}}
\newcommand{\ea}{\end{array}}
\usepackage[dvips]{epsfig}

\begin{document}

\title[Generalized nonpolynomial Schr\"odinger equations]
{Generalized nonpolynomial Schr\"odinger equations \\ 
for matter waves under anisotropic transverse confinement}
\author{Luca Salasnich}
\address{CNR-INFM and CNISM, Dipartimento di Fisica ``Galileo Galilei'', 
Universit\`a di Padova, Via Marzolo 8, 35131 Padova, Italy} 

\ead{salasnich@pd.infn.it}

\begin{abstract} 
Starting from the three-dimensional Gross-Pitaevskii equation (3D GPE)  
we derive a 1D generalized nonpolynomial Schr\"odinger 
equation (1D g-NPSE) which describes the dynamics of 
Bose-Einstein condensates 
under the action of a generic potential in the longitudinal 
axial direction and of an anisotropic harmonic potential 
in the transverse radial direction. This equation reduces to the familiar 
1D NPSE [Phys. Rev. A {\bf 65}, 043614 (2002)] in the case of isotropic 
transverse harmonic confinement. In addition we show that if the longitudinal 
potential models a periodic optical lattice the 3D GPE can be mapped 
into a 1D generalized discrete nonpolynomial Schr\"odinger 
equation (1D g-DNPSE). 
\end{abstract}


\maketitle

\section{Introduction}

At ultralow temperature dilute Bose-Einstein condensates (BECs) 
can be accurately described by the three-dimensional Gross-Pitaevskii 
equation (3D GPE) \cite{leggett}. Some years ago we have found 
that, starting from the 3D GPE 
and using a Gaussian variational approach \cite{PerezGarcia98,kav}, 
it is possible to derive an effective 
1D wave equation that describes the axial dynamics of a Bose condensate 
confined in an external potential with cylindrical symmetry 
\cite{sala1,sala2}. In this derivation the trapping potential is 
harmonic and isotropic in the transverse direction 
and generic in the axial one. Our equation, that is a time-dependent 
nonpolynomial nonlinear Schr\"odinger equation (1D NPSE) \cite{sala1,sala2}, 
has been used to model cigar-shaped condensates by many experimental 
and theoretical groups (see for instance \cite{toscani,russi,tedeschi,
spagnoli}). Moreover, by using 
the 1D NPSE we have found analytical and numerical solutions 
for solitons and vortices, which generalize the ones known 
in the literature \cite{sala3}. Recently, we have 
obtained a disrete version of NPSE, the 1D DNPSE, which 
models Bose-Einstein condensate confined in a combination 
of a cigar-shaped trap and deep optical lattice acting in 
the axial direction \cite{sala4}. 

A relevant limitation of the 1D NPSE \cite{sala1,sala2} is 
the fact that the transverse harmonic confinement is isotropic. 
To overcome this problem in this paper we introduce a 
1D generalized nonpolynomial Schr\"odinger 
equation (1D g-NPSE) which describes the dynamics of 
matter waves under the action of a generic potential in the longitudinal 
axial direction and of an anisotropic harmonic potential 
in the transverse radial direction. This equation reduces to the familiar 
1D NPSE \cite{sala1,sala2} in the case of isotropic 
transverse harmonic confinement. In the second part of the paper 
we consider a deep periodic optical lattice in the axial direction. 
In this case we find that the 3D GPE can be mapped 
into a 1D generalized discrete nonpolynomial Schr\"odinger 
equation (1D g-DNPSE), which becomes the 1D DNPSE obtained 
in Ref. \cite{sala4} if the matter waves are confined by a transverse 
isotropic harmonic trap. 

\section{1D generalized NPSE}

We consider a dilute BEC confined in the axial direction by a
generic potential $V(z)$ and in the transverse plane
by the anisotropic harmonic potential
\beq
U(x,y) = {m\over 2} \omega_{\bot}^2
\left( \lambda^2 x^2 + y^2 \right) \; ,
\eeq
where $\lambda$ is the
anisotropy of the transverse harmonic trap:
$\omega_x=\lambda \omega_{\bot}$
and $\omega_y=\omega_{\bot}$ are the two harmonic
trapping frequencies in the $x$ and $y$ directions.
The characteristic harmonic length is 
$a_{\bot}=(\hbar/(m\omega_{\bot}))^{1/2}$
and so the characteristic length and time units
are $a_{\bot}$ and $\omega_{\bot}^{-1}$,
and the characteristic energy unit is $\hbar \omega_{\bot }$.
The system is well described by the 3D GPE, and
in scaled units it reads
\beq
i{\frac{\partial \psi }{\partial t}} =
\left[ -{\frac{\hbar^2}{2m}}\nabla^{2}
+ {1\over 2} \left( \lambda^2 x^2 + y^2 \right)
+ V(z) +2\pi g |\psi |^{2}\right] \psi \; , 
\label{3dgpe}
\eeq
where $\psi(\mathbf{r},t)$ is the macroscopic wave function of the
condensate normalized to unity, and $g= 2Na_s/a_{\bot}$, with $N$ the
number of atoms and $a_s$ the s-wave scattering length of the
inter-atomic potential.
This equation can be derived from the Lagrangian density
\beq
\mathcal{L} =
{i\over 2} \big( \psi^{\ast } \frac{\partial \psi }{\partial t} 
+ \psi \frac{\partial \psi^{\ast} }{\partial t} \big) 
-{\frac{1}{2}}|\nabla \psi |^{2}
-{\frac{1}{2}}(\lambda^2 x^2 + y^2) |\psi |^2 
-V(z)|\psi |^2 - \pi g|\psi |^4 \;.
\label{lagrangian}
\eeq

To reduce the 3D problem to a 1D one, we apply a variational approach. 
In the case of an anisotropic cigar-shaped geometry, it is natural
to extend the variational representation \cite{sala1,sala2} of the 3D GPE
which led to the NPSE for the axial wave function in the
isotropic case to the present situation.
Thus we adopt the following ansatz for
the state described by 3D GPE
\beq
\psi
= {1\over \sqrt{\pi} \sigma \eta}
\exp{\left[ - \left( {x^2\over 2\sigma^2}
+ {y^2\over 2\eta^2} \right) \right] }\,f \; ,
\eeq
where $\sigma(z,t)$, $\eta(z,t)$ and $f(z,t)$, which account
for transverse widths and axial wave function,
are the 3 effective fields to be determined variationally.

By inserting this ansatz into the Lagrangian density (\ref{lagrangian}),
performing the integration over $x$ and $y$ and neglecting spatial 
derivatives of transverse widths (adiabatic approximation \cite{sala1}),
we derive the effective Lagrangian density
\beqa
\bar{\mathcal{L}} &=&
{i\over 2} \big( f^{\ast }\frac{\partial f}{\partial t} 
+ f\frac{\partial f^{\ast}}{\partial t} \big) 
- \frac{1}{2}
\left\vert{\frac{\partial f}{\partial z}}\right\vert^{2} 
- \frac{1}{4} \left( \frac{1}{\sigma^2} +
\frac{1}{\eta^2} \right) |f|^2
\nonumber 
\\
&&- {1\over 4}
\left( \lambda^2 \sigma^2 + \eta^2 \right) |f|^2
- V(z)|f|^2 - \frac{1}{2} g \frac{|f|^4}{\sigma \eta} \; .
\label{effective}
\eeqa
This effective Lagrangian gives rise to a system of 3 Euler-Lagrange
equations, obtained by varying $\bar{\mathcal{L}}$ with respect to
$f^{\ast }$, $\sigma$ and $\eta$:
\beqa
i \frac{\partial f}{\partial t} &=&
\Big[ - \frac{1}{2} \frac{\partial^2}{\partial z^2} + 
\frac{1}{4} \left( \frac{1}{\sigma^2} +
\frac{1}{\eta^2} + \lambda^2 \sigma^2 + \eta^2 \right) 
+ V(z) + g \frac{|f|^2}{\sigma \eta} \Big] f \; ,
\label{d-npse}
\\
\lambda^2 \sigma^4 &=& 1 + g |f|^2 {\sigma\over \eta} \; ,
\quad\quad\quad\quad\quad\quad\quad
\;
\label{sigma1}
\\
\eta^4 &=& 1 + g |f|^2 {\eta\over \sigma} \; .
\quad\quad\quad\quad\quad\quad\quad
\label{sigma2}
\eeqa
We call this system of three coupled equations,
which describe the BEC under a transverse anisotropic
harmonic confinement, the 1D generalized nonpolynomial
Schr\"odinger equation (1D g-NPSE). Notice that
with $g\neq 0$ only for $\lambda=1$ one finds
$\sigma=\eta$. In this case of isotropic 
transverse harmonic confinement ($\lambda=1$) we have 
\beq 
\sigma^4 = \eta^4 =  1 + g |f|^2 \; , 
\eeq 
and the g-NPSE becomes  
\beq 
i {\partial\over \partial t} f = 
\Big[ - \frac{1}{2} \frac{\partial^2}{\partial z^2} + V(z) \Big] f 
+ {1+(3/2) g|f|^2\over \sqrt{1+g|f|^2}} f \; ,  
\eeq
which is the 1D NPSE derived in Ref. \cite{sala1,sala2}. 
Instead, for an anisotropic transverse harmonic confinement 
($\lambda\neq 1$) the transverse widths $\sigma$ 
and $\eta$ depend on $g$ in a not trivial way.

\subsection{Weak-coupling regime}

In the weak-coupling regime, i.e. $g|f|^2 \ll 1$, 
we can expand in series of powers of $g|f|^2$ the widths $\sigma$
and $\eta$. At the first order in $g|f|^2$, the g-NPSE gives
\beqa
\label{perturbationw}
&&\sigma=\frac{1}{\sqrt{\lambda}}-\frac{1}{4 \lambda}g|f|^2
\\
&&\eta=1-\frac{\sqrt{\lambda}}{4}g|f|^2
\eeqa
and we get a 1D GPE for the field $f(z,t)$
\beq
i \frac{\partial f}{\partial t} = \Big[ -
\frac{1}{2} \frac{\partial^2}{\partial z^2} +\frac{1}{2}
\left(\lambda + 1 \right) + V(z) + g\sqrt{\lambda} |f|^2 
\Big] f.
\label{1dgpe}
\eeq
In this weak-coupling limit the anisotropy produces a renormalization
$g\sqrt{\lambda}$ of the interaction strength $g$ and the
transverse energy is simply $(\lambda +1)/2$.

\subsection{Strong-coupling regime}

In the strong-coupling regime $g|f|^2 \gg 1$ the g-NPSE gives
\beqa 
\sigma&=& {(g|f|^2)^{1/4}\over \lambda^{3/4}} 
\\
\eta&=&\lambda^{1/4} (g|f|^2)^{1/4}
\eeqa
and the differential equation reads 
\beq
i \frac{\partial f}{\partial t} = \Big[ -
\frac{1}{2} \frac{\partial^2}{\partial z^2} + V(z)|f|^2 + {3\over
2} \sqrt{g\lambda} |f| \Big] f \; ,
\eeq
that is a 1D Schr\"odinger equation with quadratic nonlinearity.
In this regime, after setting
\beq
f(z,t)=\rho_1(z)^{1/2} \exp{(- i \mu t)} \ ,
\eeq
where $\mu$ is the adimensional chemical potential, and neglecting
the spatial derivative (Thomas-Fermi approximation), one finds the
stationary axial density profile of the BEC
\beq
\rho_1(z) = {4\over 9 g\lambda} \left( \mu - V(z) \right)^2 \; .
\eeq

\subsection{Quasi-2D regime}

Under the condition $g|f|^2\eta \ll \sigma$, which implies a 
very strong harmonic confinement along the $y$ axis ($\lambda \ll 1$), 
Eq. (\ref{sigma2}) gives $\eta \simeq 1$ and the system is quasi-2D. 
In this case from Eqs. (\ref{d-npse}) and (\ref{sigma1}) we get 
\beq
i {\frac{\partial}{\partial t}} f = \left[ - {\frac{1}{2}}{\frac{\partial^2}
{\partial z^2}} + V(z) + {\frac{1}{4}} 
\left( {\frac{1}{\sigma^2}} + \lambda^2 \sigma^2 +2 \right) \right] f 
+ {\frac{g}{\sigma}} |f|^2 f \; ,  
\label{1dnpse}
\eeq
and also 
\beq
\lambda^2 \sigma^4 = 1 + g |f|^2 \sigma \; .  
\label{sigma}
\eeq
We observe that this last equation, based on a 3D$\to$2D$\to$1D reduction 
of the GPE, is formally equivalent (with $\lambda=1$) to an equation 
for the width $\sigma$ previously obtained within a 3D$\to$2D reduction 
of the GPE \cite{2dnpse}. Exact solutions to Eq. (\ref{sigma}) 
are given by the Cardano formula,
\beqa
\sigma 
&=& {\frac{1}{\lambda^{1/2}}} \Big\{ \pm {\frac{1}{2}}\sqrt{\frac{A^2 -12
}{3A}} 
\\ 
&+&{\frac{1}{2}}\sqrt{-{\frac{A^2-12}{3A}}\pm 2{\frac{g}{
\lambda^{1/2}}} |f|^2 \left( \frac{A^2-12}{3A}\right)^{-1/2}} \Big\} \; ,
\label{eta-solve}
\nonumber
\eeqa
where the upper and lower signs correspond, respectively, to $g>0$ and 
$g<0$, and
\beq
A= \left( {3/2}\right) ^{1/3}\left( {\frac{9g^2}{\lambda}} |f|^4
+ \sqrt{3} \sqrt{256+ {\frac{27g^4}{\lambda^2}} |f|^8}\right)^{1/3} \; .
\label{A}
\eeq

\subsection{Analytical-numerical approach}

In general, the solutions of Eqs. (\ref{sigma1}) and (\ref{sigma2})
do not have a simple analytical expression. Nevertheless
from Eqs. (\ref{sigma1}) and (\ref{sigma2}) one finds that
the transverse widths $\sigma$ and $\eta$ can be expressed
in parametric form as
\beqa
\sigma &=& g|f|^2 {(\xi+1)^{1/4}\over \xi} \; ,
\label{pip1}
\\
\eta &=& (\xi+1)^{1/4} 
\label{pip2}
\eeqa 
where the parameter $\xi$ is the 
solution of the quartic equation 
\beq 
\xi^3 \ (\xi + g^2|f|^4 ) = \lambda^2 g^4 |f|^8 \ (\xi + 1) \; .   
\label{alg-eq}
\eeq 
The algebraic equation (\ref{alg-eq}) can be solved numerically. 
Nevertheless some analytical results can be easily obtained 
in limiting case. Eqs. (\ref{pip1}) and (\ref{pip2}) show that for a
repulsive BEC ($g>0$) it must be $\xi >0$, while for an attractive BEC
($g<0$) it is $-1 \le \xi < 0$. 
If $\lambda=1$ then $\xi=g|f|^2$ 
and $\sigma=\eta=(1+g|f|^2)^{1/4}$. 
If $g|f|^2=1$ then $\xi=\lambda^{2/3}$ and $\sigma
=\eta/\lambda^{2/3}$ with $\eta=(1+\lambda^{2/3})^{1/4}$. 
If $g|f|^2=-1$ then $\xi=-1$ and $\sigma=\eta=0$. 

\begin{figure}
\begin{center}
{\includegraphics[width=8.cm,clip]{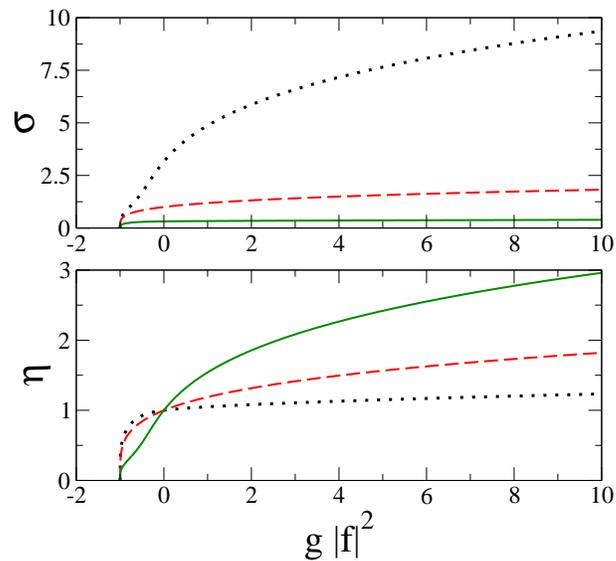}}
\caption{Transverse widths $\sigma$ 
and $\eta$ as a function of the axial 
strength $g|f|^2$. Three values of the transverse anisotropy: 
$\lambda=0.1$ (dotted curves); $\lambda=1$ (dashed curves); 
$\lambda=10$ (solid curves).} 
\label{fig-sig1sig2}
\end{center}
\end{figure}

In Fig. \ref{fig-sig1sig2} we report the transverse widths $\sigma$ 
and $\eta$ obtained from Eqs. (\ref{pip1}), (\ref{pip2}) 
and the numerical solution of Eq. (\ref{alg-eq}) with the 
Newton-Raphson method. We plot the widths as a function of the 
axial strength $g|f|^2$ for three values of the transverse 
anisotropy $\lambda$. 

\begin{figure}
\begin{center}
{\includegraphics[width=8.cm,clip]{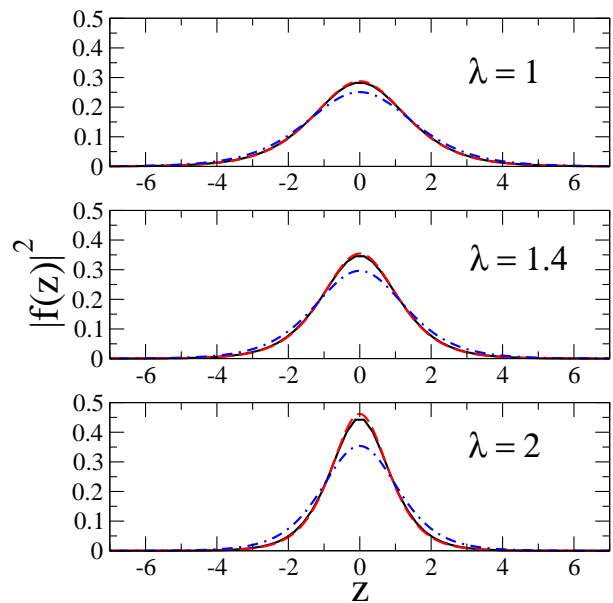}}
\caption{Axial density profile $|f(z)|^2$ 
of the triaxial bright soliton, i.e. attractive BEC without 
axial confinement ($V(x)=0$). 
Solid curves: numerical solutions of the 3D GPE, Eq. (\ref{3dgpe}). 
Dashed curves: numerical solutions of the 1D g-NPSE, Eq. (\ref{d-npse}).  
Dot-dashed curves: numerical solutions of the 1D GPE, Eq. (\ref{1dgpe}). 
Interaction strength $g=-1$ and three values 
of the transverse anisotropy $\lambda$.} 
\label{fig-compare}
\end{center}
\end{figure}

To test the accuracy of 1D g-NPSE we consider the case with 
negative scattering length ($g<0$) without axial confinement 
($V(x)=0$). Under these conditions the system admits 
triaxial bright soliton configurations \cite{sala-triaxial}. 
In Fig. \ref{fig-compare} we plot the axial density profile $|f(z)|^2$ 
of the BEC with $g=-1$ and three values of the transverse anisotropy $\lambda$.
The solid curves represent the numerical 
results obtained by solving the 3D GPE, Eq. (\ref{3dgpe}), 
with a finite-difference 
Crank-Nicolson algorithm with imaginary time \cite{sala-numerics}. 
In this case the axial density profile is given by 
\beq 
\rho(z) = \int |\psi({\bf r})|^2\ dx \ dy \; ,  
\eeq
with $\psi({\bf r})$ ground-state wave function of the 3D GPE. 
The dashed curves are obtained by solving the g-NPSE and 
the dot-dashed curves by the 1D GPE. 
The figure shows that the results of g-NPSE are in good 
agreement with the 3D GPE ones, while the predictions of the 
1D GPE, Eq. (\ref{1dgpe}), are not very reliable for the 
interaction strength $g=-1$.  

\section{1D generalized DNPSE} 

Let us consider again Eq. (\ref{3dgpe}) where now the 
axial potential is given by 
\beq 
V(z) = V_0 \cos{(2 k z)} \; .  
\eeq
This potential models the optical lattice produced in experiments with 
Bose-Einstein condensates by using counter-propagating laser beams 
\cite{oliver}. 

To symplify the problem we set 
\beq
\psi(x,y,z,t) = \sum_j \phi_j(x,y,t) \ W_j(z) 
\eeq
where $W_j(z)$ is the Wannier function maximally localized 
at the $j$-th minimum of the axial periodic potential. 
This tight-binding ansatz is reliable in the case of a deep optical 
lattice \cite{oliver,tromba}. 
We insert this ansatz into Eq. (\ref{3dgpe}), multiply 
the resulting equation by $W_n^*(z)$ and integrate over $z$ variable. 
In this way we get 
\beq
i{\partial\over \partial t} \phi_n = 
\left[-\frac{1}{2}\nabla_{\bot}^2 + 
\frac{1}{2} \left( \lambda^2 x^{2}+y^{2}\right) 
+ \epsilon \right] \phi_n 
- J \left( \phi_{n+1}+\phi_{n-1}\right) + 
2\pi \gamma \left\vert \phi_n \right\vert^{2} \phi_n 
\: , 
\label{array2D}
\eeq
where the parameters $\epsilon$, $J$ and $\gamma$ are given by 
\beq
\epsilon = \int W_n^*(z) \left[ -{1\over 2} {\partial^2\over \partial z^2} 
+ V_{0}\cos \left( 2kz\right) \right] W_n(z) \ dz  \; , 
\eeq
\beq
J = - \int W_{n+1}^*(z) \left[ -{1\over 2}
{\partial^2\over \partial z^2} 
+  V_{0}\cos \left( 2kz\right) \right] W_n(z) \ dz  \; , 
\eeq
\beq 
\gamma = g \int |W_n(z)|^4 \ dz  \; . 
\eeq
In the tight-binding regime the parameter $J$ is positive definite. 
In addition the parameters $\epsilon$, $J$ and $\gamma$ are independent 
on the site index $n$. 

Eq. (\ref{array2D}) can be seen as the Euler-Lagrange equation 
of the following Lagrangian density
\beqa
{\cal L} &=& \sum_n \Big\{ 
\phi_n^* \left[ i {\partial\over \partial t} + 
{1\over 2} \nabla_{\bot}^2 - {1\over 2} (\lambda^2 x^2+y^2) 
-\epsilon \right] \phi_n 
\nonumber
\\
&&
+ J \phi_n^* \left( \phi_{n+1}+\phi_{n-1} \right) - \pi \gamma |\phi_n|^4 
\Big\} \; .  
\label{lagr}
\eeqa

To further simplify the problem we set 
\beq
\phi_n = 
{1\over \sqrt{\pi} \sigma_n \eta_n}
\exp{\left[ - \left( {x^2\over 2\sigma_n^2}
+ {y^2\over 2\eta_n^2} \right) \right] }\,f_n \; , 
\label{assume}
\eeq
where $\sigma_n(t)$, $\eta_n(t)$ and $f_n(t)$, which account
for discrete transverse widths and axial wave function, 
are the effective generalized coordinates to be determined variationally. 

We insert this ansatz into the Lagrangian density (\ref{lagr}) 
and integrate over $x$ and $y$ variables. 
In this way we obtain an effective Lagrangian for the fields 
$f_n(t)$ and $\sigma_n(t)$. This Lagrangian reads 
\beqa
L &=& \sum_n \Big\{ f_n^* \left[ i {\partial\over \partial t} - 
{1\over 4} \left( {1\over \sigma_n^2} + \lambda^2 \sigma_n^2 
+ {1\over \eta_n^2} + \eta_n^2 \right) 
-\epsilon \right] f_n 
\nonumber 
\\
&&
+ J f_n^* \left( f_{n+1}+f_{n-1} \right) - 
{\gamma \over 2\sigma_n \eta_n} |f_n|^4 \Big\} \; . 
\label{Lagrangian}
\eeqa
To obtain this expression we have supposed that the transverse widths 
$\sigma_n(t)$ of nearest-neighbor sites are practically equal 
(this is the equivalent to the adiabatic approximation made 
for the continuous model). The Euler-Lagrange equation of (\ref{Lagrangian}) 
with respect to $f_n^*$ is 
\beqa 
i {\partial\over \partial t} f_n &=& 
\left[ {1\over 4} \left( {1\over \sigma_n^2} + \lambda^2 
\sigma_n^2 + {1\over \eta_n^2} + \eta_n^2 \right) +\epsilon \right] f_n 
\nonumber
\\
&&
- J \left( f_{n+1}+f_{n-1} \right) + 
{\gamma \over \sigma_n \eta_n} |f_n|^2 f_n \; . 
\label{e1}
\eeqa
while the Euler-Lagrange equations of (\ref{Lagrangian}) 
with respect to $\sigma_n$ and $\eta_n$ give 
\beqa 
\lambda^2 \sigma_n^4 = 1 + \gamma |f_n|^2 {\sigma_n\over \eta_n} \; ,
\label{e1-sig}
\\
\eta_n^4 = 1 + \gamma |f_n|^2 {\eta_n\over \sigma_n} \; .
\label{e1-eta}
\eeqa
We call this system of coupled equations,
which describe the BEC under a transverse anisotropic
harmonic confinement and an axial optical lattice, 
the 1D generalized and discrete nonpolynomial 
Schr\"odinger equation NPSE (1D g-DNPSE)
In the case of isotropic transverse harmonic confinement 
($\lambda=1$) we have 
\beq 
\sigma_n^4 = \eta_n^4 =  1 + \gamma |f_n|^2 \; , 
\eeq 
and the 1D g-DNPSE becomes  
\beq 
i {\partial\over \partial t} f_n = \epsilon f_n 
- J \left( f_{n+1}+f_{n-1} \right) + 
{1+(3/2) \gamma |f_n|^2\over \sqrt{1+g|f_n|^2}} f_n \; ,  
\eeq
which is substantially the 1D DNPSE derived in Ref. \cite{sala4}. 

The 1D g-DNPSE given by Eqs. (\ref{e1}), (\ref{e1-sig}) and (\ref{e1-eta}) 
is the discrete version of 1D g-NPSE given by Eqs. (\ref{d-npse}), 
(\ref{sigma1}) and (\ref{sigma2}). It is then strightforward 
to repeat with the 1D g-DNPSE the analysis of weak-coupling, 
strong-coupling and quasi-2D regimes we have performed 
with the 1D g-NPSE. 

\section{Conclusions} 

We have obtained new nonpolynomial Schr\"odinger 
equations for matter waves under anisotropic transverse 
harmonic confinement. In the case of isotropic 
transverse confinement these continuous and discrete equations 
reduce to the ones derived in previous papers. 
The generalized 1D continuous nonpolynomial Schr\"odinger 
equation can be used to study the static and dynamics of 
Bose condensates with a generic longitudinal axial potential. 
The generalized 1D discrete nonpolynomial Schr\"odinger 
equation can be insead used to investigate the effect of 
a deep axial periodic potential. The comparison 
with the full 3D Gross-Pitaevskii equations show that 
these effective 1D equations are indeed quite accurate 
and much simpler for numerical and analytical computations. 

The author thanks Alberto Cetoli, Boris Malomed, Giovanni Mazzarella, 
Dmitry Pelinovsky and Flavio Toigo for useful comments and suggestions. 
This work has been partially supported by Fondazione CARIPARO.


\section*{References}
\begin{thebibliography}{10}

\bibitem{leggett} A.J. Leggett, 
{\it Quantum Liquids. Bose Condensation and Cooper Pairing in 
Condensed-Matter Systems} (Oxford Univ. Press, Oxford, 2006). 

\bibitem{PerezGarcia98} V. M. P\'{e}rez-Garc\'{\i}a, H. Michinel, and H.
Herrero, Phys. Rev. A \textbf{57}, 3837 (1998).

\bibitem{kav} A.D. Jackson, G.M. Kavoulakis, 
and C.J. Pethick, Phys. Rev. A 58, 2417 (1998). 

\bibitem{sala1} L. Salasnich, A. Parola, and L. Reatto, 
Phys. Rev. A \textbf{65}, 043614 (2002).

\bibitem{sala2} L. Salasnich, Laser Phys. \textbf{12}, 198 (2002). 

\bibitem{toscani} P. Massignan and M. Modugno, 
Phys. Rev. A {\bf 67}, 023614 (2003);  
M. Modugno, C. Tozzo, and F. Dalfovo, 
Phys. Rev. A {\bf 70}, 043625 (2004); 
C. Tozzo, M. Kramer, and F. Dalfovo, 
Phys. Rev. A {\bf 72} 023613 (2005): 
M Modugno, Phys. Rev. A {\bf 73}, 013606 (2006). 

\bibitem{russi} A.M. Kamchatnov and V.S. Shchesnovich, Phys. Rev. A 
{\bf 70} 023604 (2004); D. E. Pelinovsky, P.G. Kevrekidis, D.J. 
Frantzeskakis, and V. Zharnitsky, 
Phys. Rev. E {\bf 70}, 047604 (2004). 

\bibitem{tedeschi} 
G. Theocharis, P. G. Kevrekidis, M. K. Oberthaler, and D. J. Frantzeskakis, 
Phys. Rev. A {\bf 76}, 045601 (2007); A. Weller, J. P. Ronzheimer, C. Gross,  
J. Esteve, M. K. Oberthaler, D. J. Frantzeskakis, G. Theocharis, 
and P.G. Kevrekidis, Phys. Rev. Lett. {\bf 101}, 130401 (2008). 

\bibitem{spagnoli}  A. Munoz Mateo and V. Delgado, 
Phys. Rev. A {\bf 75}, 063610 (2007); 
A. Munoz Mateo and V. Delgado, 
Phys. Rev. A {\bf 77}, 013617 (2008); 
A. Munoz Mateo and V. Delgado, Ann. Phys. {\bf 324}, 709 (2009). 

\bibitem{sala3} L. Salasnich, A. Parola, and L. Reatto, Phys. Rev. A 
\textbf{66}, 043603 (2002); L. Salasnich, Phys. Rev. A {\bf 70}, 
053617 (2004); L. Salasnich, A. Parola and L. Reatto, 
Phys. Rev. A {\bf 70}  013606 (2004); 
A. Parola, L. Salasnich, R. Rota, and L. Reatto, 
Phys. Rev. A {\bf 72}, 063612 (2005); L. Salasnich and B.A. Malomed, 
Phys. Rev. A {\bf 74}, 053610 (2006). 

\bibitem{sala4} A. Maluckov, L. Hadzievski, B.A. Malomed, and L. Salasnich, 
Phys. Rev. A \textbf{78}, 013616 (2008). 

\bibitem{2dnpse} L. Salasnich and B.A.
Malomed, \textit{Solitons and solitary vortices in ``pancake"-shaped
Bose-Einstein condensates}, submitted for publication (2009).

\bibitem{sala-triaxial} L. Salasnich, Laser Phys. {\bf 15}, 366 (2005). 

\bibitem{sala-numerics} E. Cerboneschi, R. Mannella, E. Arimondo, and L.
Salasnich, Phys. Lett. A \textbf{249}, 495 (1998); L. Salasnich, A. Parola,
and L. Reatto, Phys. Rev. A \textbf{64}, 023601 (2001).

\bibitem{oliver} O. Morsch and M. Oberthaler, 
Rev. Mod. Phys. {\bf 78}, 179 (2006).  

\bibitem{tromba} A. Trombettoni and A. Smerzi, Phys. Rev. Lett. {\bf 86}, 
2353 (2001); A. Smerzi and A. Trombettoni, P.G. Kevrekidis, and 
A.R. Bishop, Phys. Rev. Lett. {\bf 89}, 170402 (2002); 
A. Smerzi and A. Trombettoni, Phys. Rev. A {\bf 68}, 
023613 (2003). 

\end{thebibliography}
\end{document}